\documentclass[aps,prb,reprint,groupedaddress,floatfix,showpacs]{revtex4-1}
\usepackage{graphicx}
\usepackage{epstopdf}
\begin{document}

\title{Classical model of negative thermal expansion in solids
with expanding bonds}


\author{Joseph T. Schick}
\email[]{joseph.schick@villanova.edu}
\affiliation{Department of Physics, Villanova University, Villanova, 
PA 19085, USA}

\author{Andrew M. Rappe}
\affiliation{The Makineni Theoretical Laboratories, Department of Chemistry, 
University of Pennsylvania, Philadelphia, Pennsylvania 19104-6323, USA}


\date{\today}

\begin{abstract}
We study negative thermal expansion (NTE) in model
lattices with multiple atoms per cell and first- and second-nearest
neighbor interactions using the (anharmonic) Morse potential.  By exploring 
the phase space of neighbor distances and thermal expansion rates of 
the bonds, we determine the conditions under which NTE emerges.  
By permitting all bond lengths to expand at different rates,
we find that NTE is possible without appealing to fully rigid units. 
Nearly constant, large-amplitude, isotropic NTE is observed up to the melting 
temperature in a classical molecular dynamics model of a $\mathrm{ReO}_3$-like 
structure when the rigidity of octahedral units is almost completely
eliminated. Only weak NTE, changing over to positive expansion
is observed when the corner-linked octahedra are rigid, with
flexible second-neighbor bonds between neighboring octahedra
permitting easy rotation.
We observe similar changes to thermal expansion behavior for the 
diamond lattice: NTE when second-neighbor interactions are weak to
positive thermal expansion when second-neighbor interactions are
strong.  From these observations, we suggest that the only essential
local conditions for NTE are atoms with low coordination numbers along
with very low energies for changing bond angles relative to 
bond-stretching energies.
\end{abstract}


\maketitle

\section{Introduction}

Negative thermal expansion (NTE) is of great technological interest
because of the potential to control or eliminate thermal stresses in
real objects and devices.
The material ZrW$_2$O$_8$ has been at the focus of intensive
study because it exhibits substantial isotropic NTE over a wide
temperature 
range.\cite{Martinek68p227,Evans96p2809,Mary96p90,Drymiotis04p025502}  
While there is an expanding literature on the microscopic causes of
NTE in this material, no model has proven completely
adequate in making this connection.
\cite{Mary96p90,Pryde96p10973,Ramirez98p4903,Ernst98p1247,Mittal99p7234,Mittal01p4692,Cao02p215902,Cao03p014303,Hancock04p225501,Tucker05p255501,Gava12p195503,Bridges14p045505,Sanson14p3716}

The picture at the heart of the discussion in ZrW$_2$O$_8$ is that 
the essential ionic motions leading to thermal changes in
the lattice dimensions can be described
by collective rotations of groups of atoms, the foundation
of the ``rigid unit model'' (RUM),\cite{Tucker05p255501} 
the rigid units being composed of WO$_4$ tetrahedra and ZrO$_6$
octahedra.
Citing the exceptional rigidity of the Zr-O-W lines in the
lattice seen in x-ray absorption fine-structure experiments,
yet indicating distortion of the polyhedra, 
Cao and coauthors created a variation of the
RUM labeled the ``tent model,'' in which lattice contraction
is a result of coherent motion of larger groups of 
atoms in the lattice.\cite{Cao02p215902,Cao03p014303}  
This disagreement in interpretation motivated  
a molecular dynamics study,\cite{Sanson14p3716} employing 
empirical potentials from previous work,\cite{Pryde96p10973} 
leading to the conclusion that the only 
essentially rigid units in the lattice are individual W-O and Zr-O 
bonds, questioning the validity of both the
RUM and tent models.\cite{Sanson14p3716}

NTE has been observed via x-ray powder diffraction\cite{Sennova07p290}
and coherent neutron powder diffraction
experiments in Li$_2$B$_4$O$_7$.\cite{Senyshyn10p093524,Senyshyn12p175305}
The expansion in this material is anisotropic, with the
contraction of one lattice parameter dominating the 
low-temperature volume dependence.  
The Li ions show strong $T$ dependence of their mean-squared
displacements around 100~K, 
temperature at which NTE vanishes in the material.
In the NTE regime, it is observed that the angle
formed by Li-Li-Li chains correlates with the
contracting lattice parameter,\cite{Sennova07p290,Senyshyn10p093524}
indicating the contribution of bond-bending to NTE.

Materials with simpler structure also exhibit NTE.
For example, ranges of NTE are experimentally observed
in diamond/zincblende structured materials Si, Ge, GaAs,
GaSb, InAs, and InSb.\cite{Sparks67p779} 
In a comparative numerical study of models for
NTE in CuO$_2$, classical potentials were used to build three
vibrational models based on rigid polyhedra, or rigid two- or 
three-atom rods.\cite{Rimmer14p214115}
The resulting phonon dispersion curves were compared to the 
\emph{ab initio} curves, with results suggesting
that the low-energy modes correspond to rigid Cu-O 
rods.\cite{Rimmer14p214115}  
In a density functional theoretical study of 
NTE in Ag$_2$O, Cu$_2$O, and Au$_2$O, the variation in
NTE across these compounds is observed to correlate with
the degree to which charge density is spherically
distributed around the atoms, again supporting the
importance of bond bending as a contributing
factor in NTE.

ReO$_3$ consists of a cubic
lattice of octahedral units centered on Re ions.
NTE is observed in the material over two temperature
ranges, from 2~K to 220~K and from 
600~K to 680~K, in an experiment that includes
temperatures up to about 800~K.\cite{Chatterji09p241902}
Experimental evidence supports ReO$_3$
maintaining its cubic phase ($Pm\bar{3}m$) down to lowest 
temperatures\cite{Bozin12p094110} and up to 800~K 
temperature.\cite{Chatterji09p241902}
It has been suggested that NTE in ReO$_3$ is readily
explained within the RUM.\cite{Chatterji09p241902}
These neutron studies indicate that O-atom motion transverse
to Re-O bonds is significant and increases with
temperature while longitudinal motion of the O atoms
is minimal and independent of temperature, as evidenced by
analysis of the mean-squared displacements of the 
ions.\cite{Chatterji09p241902} 
While this evidence supports treating ReO$_3$ octahedral units as 
rigid, attributing the contraction of the lattice to
coherent rotations of the octahedra, it does not
eliminate the possibility of octahedral distortions
that preserve the Re-O distances.
Lending support to this view are the results of 
experimental and theoretical work reported on isostructural
ScF$_3$, in which the strong NTE observed is explained
partly in terms of the distortions of the ScF$_6$
octahedra.\cite{Li11p195504}  Additionally, the expansion of 
the Sc-F bonds with temperature is demonstrated in a recent
density-functional molecular dynamics study that produces
thermal expansion of ScF$_3$, comparing well with
experimental results.\cite{Lazar15p224302}

All vibration in materials is based on the fact that atoms
sit in potential wells determined by 
their interactions with their neighbors, \emph{i.e.}\ 
through the intervening chemical bonds.  As the temperature
is increased, the nuclei
vibrate with greater amplitude.
Because of the strong repulsion between atoms at close range,
the potential between atoms is anharmonic.  Thermal vibrations
of atoms interacting through such an anharmonic potential 
generally drive neighboring 
atoms farther apart as temperature increases.
In solids, the expansion of bonds is evident as thermal
expansion.  So how can contraction arise?

Especially as noted above in the discussion
of modeling NTE in ZrW$_2$O$_8$,
no simple model proves to be universally
capable of explaining all NTE behavior, leading us to consider
that by resolving a material into
these units, some essential physics may have been neglected.
Perhaps another approach may shed light on the situation.
One aspect common to any rigid unit related model is the
assumption that expansion of some bonds can
be neglected, while molecular dynamics calculations
indicate that more flexibility is necessary.\cite{Sanson14p3716}  
A model without the rigidity requirement may help connect
the influence of all-bond expansion to NTE.

Theoretical investigations of the general features of
NTE include an analytical calculation of rigid corner 
linked squares\cite{Simon01p1781} and of 
rigid squares and octahedra joined together 
with flexible links,\cite{He10p014111} demonstrating
that such models can exhibit NTE as well as
elastic constants that decrease with temperature
and pressure.
In an alternative model, Rechtsman \emph{et al.}\ 
devise an isotropic interaction potential that
leads to NTE at low temperatures in close-packed structures
of single-component many-particle systems, an effect demonstrated
through constant-pressure Monte Carlo simulations.\cite{Rechtsman07p12816}
Their designed potential rises rapidly as the components
(or atoms) move farther apart and less rapidly as they move
closer together, in contrast to the commonly-accepted
negative anharmonicity.

In this paper we present an investigation of model materials
consisting of atoms all linked by bonds of variable 
flexibility.  The flexibility is implemented 
through the Morse potential,\cite{supplemental} providing
known thermal expansion between atoms.  
(We note that NTE has been documented
in an early lattice dynamics study employing several
pair potentials, including the Morse potential.\cite{Wallace65pA152})
We show how
the ratios of interatomic equilibrium lengths and their
temperature dependences control the presence and
extent of NTE.  We examine three-dimensional models
that are designed to illuminate the physics; 
we explore the relationships 
to existing simpler analytic models, 
the effect of enhanced flexibility
of all bonds, and relationships to real NTE materials.

\section{Models}

\begin{figure}
 \includegraphics[width=2.8in]{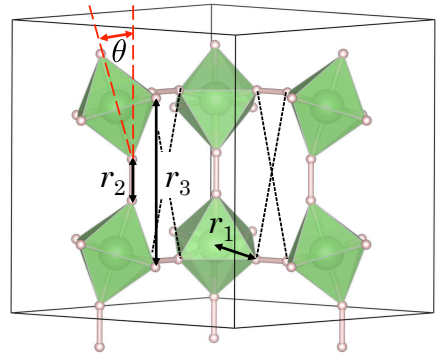}%
 \caption{\label{figure:3dmodel} (Color online) One
 unit cell of the three-dimensional model structure contains
 eight octahedra, connected by adjustable $r_2$ links,
 and pairs of second-neighbor links to provide
 the means to rotate the octahedra.  A staggered arrangement
 of the second-neighbor links enables rotation of the rigid octahedra.
 The lengths of bonds between
 the center and vertex of an octahedron is $r_1$, $r_2$ is the
 distance of the link between octahedra, and $r_3$ is the
 distance between the pairs of parallel links being 
 pushed apart by the second-neighbor interactions
 indicated with dashed lines. 
 Only two pairs of second-neighbor links are shown
 in order to enhance clarity.
 }
\end{figure}

We employ a three-dimensional model (Fig.~\ref{figure:3dmodel}) 
that bears some resemblance to previous two- and three-dimensional
work,\cite{Cao03p014303,Simon01p1781,He10p014111} however
we replace fully rigid units with seven-atom octahedra.
The octahedra form a cubic lattice,
with interactions (bonds) between pairs of atoms 
represented by Morse potentials,\cite{supplemental}
which exhibit thermal expansion.  Each now-flexible octahedron
possesses interactions between the central atom ($A$) and each
of the six vertex atoms ($X$).  The rigidity of the
octahedra is controlled by interactions between each $X$
atom and the four nearest $X$ atom vertices on the same octahedron.  
Second-neighbor interactions between $X$ atoms of different
octahedra provide bond-bending stiffness, stabilizing the 
lattice.
The center-to-vertex distance of an octahedron is
denoted $r_1$. The distances between the vertices of 
nearest-neighbor octahedra are denoted as $r_2$, and the second-
neighbor bonds as dashed lines.  We specify the 
second-neighbor distance in terms of its projection $r_3$ onto the 
nearest Cartesian axis.
When we perform calculations for corner-connected octahedra, 
$r_2=0$, the projection $r_3$ coincides with the second-neighbor
interactions, and atoms are removed from the calculation
to avoid doubled atoms at the vertices.

\subsection{Rigid mechanical model}
\label{subsection:mechanical}

\begin{figure}
 \includegraphics[width=3.2in]{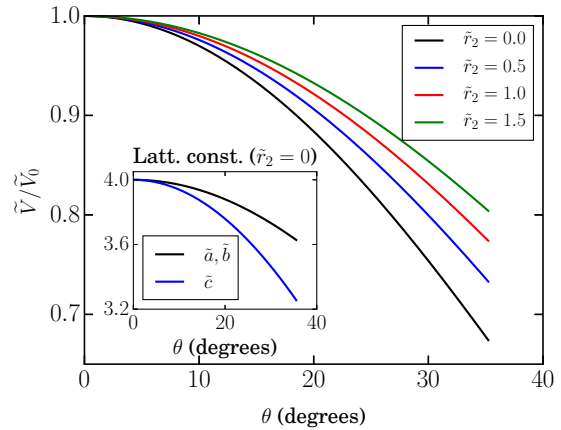}%
 \caption{\label{figure:3dvolume} 
 (Color online)
 The relative volume of the three-dimensional model 
 with tetragonal distortions due to octahedral rotations.  
 The maximum rotation angle ($\approx 35.3^\circ$) corresponds
 to alignment of parallel $\tilde{r}_2$ and $\tilde{r}_3$
 links, rendering further octahedral rotation impossible.
 The magnitude of volume contraction is maximal when the
 octahedra are corner-connected.  The contraction for
 small angle rotation is vanishingly small.
 See Fig.~\ref{figure:3dmodel} for a depiction of 
 the angle.  Inset: Normalized lattice constants of
 the corner-connected-eight-octahedra unit cell showing its
 tetragonal distortion for non-zero octahedral rotations
 are visible.}
\end{figure}

We analytically examine the effects of permitting expansion of the
second-neighbor bonds only, treating them as expanding
rods that rotate rigidly-linked and rigid octahedra.
Rotation would not be permitted with these rigid rods if they
coincide with the bonds defined above because the symmetrically 
arranged second-neighbor links between octahedra would compete
with each other.  
To permit coherent rotation of the octahedra, we temporarily
eliminate all but selected second-neighbor interactions.  
One set of these second-neighbor interactions lies in the $xz$-plane, 
with the long direction along the $z$-axis 
(see Fig.~\ref{figure:3dmodel}). 
Two additional similar sets of second-neighbor interactions lie
in the $yz$-plane with a long $z$-axis and in the 
$xy$-plane with a long $x$-axis.  Because of the asymmetry of
this arrangement, the rotation pattern represents a tetragonal 
distortion of the unit cell (space group $I4/mmm$).
The contraction of the lattice parameters are functions
of the angle $\theta$ describing the 
coherent rotations of octahedra about axes along 
alternating $\langle 110 \rangle$ directions.
To simplify the expressions below,  
we define reduced parameters $\tilde{r}_2 = r_2/r_1$ and 
$\tilde{r}_3 = r_3/r_1$.
The relation between $\theta$ and bond lengths, 
normalized to $r_1$, is
\begin{equation}
\tilde{r}_3 = \tilde{r}_2 + \sqrt{6} \cos
(\theta - \theta_\mathrm{max})\, , 
\end{equation}
where $\theta_\mathrm{max} = \tan^{-1} (\sqrt{2}/{2})
\approx 35.26^\circ$ is the angle for which
$r_2$ bonds lie in planes formed by 
parallel second-neighbor bonds.
In this configuration, 
there will be no torque produced by further expansion of 
second-neighbor bonds.  If the second-neighbor
bonds are to continue expanding, the octahedra and/or 
$r_2$ bonds would be required to distort or expand.
(The limits on $\tilde{r}_3$ are 
$2 + \tilde{r}_2 \le \tilde{r}_3 \le \sqrt{6} 
+ \tilde{r}_2$.  Reducing $\tilde{r}_3$ to less than
$2+\tilde{r}_2$ simply
reverses the sense of rotation of the octahedra, also
yielding a contraction of the lattice.)
The unit cell parameters in terms of the rotation, also 
normalized to $r_1$, are
\begin{equation}
\tilde{a} = \tilde{b} = 
2 \left( \cos \theta 
+ \tilde{r}_2  + 1 \right)
\end{equation}
and
\begin{equation}
\tilde{c} =
2 \left( 2 \cos \theta + \tilde{r}_2 \right) \, .
\end{equation}
The dimensionless volume of the unit cell within the
physically allowed limits is
\begin{equation}
\label{eq:3dvolume}
\widetilde{V}  = 8(\cos \theta 
+ \tilde{r}_2 + 1)^2 
  (2\cos \theta + \tilde{r}_2)\, .
\end{equation}
In Fig.~\ref{figure:3dvolume} we plot 
$\widetilde{V}/\widetilde{V}_\mathrm{0}$ as a function
the rotation $\theta$ 
for several values of $\tilde{r}_2$.
Because the octahedra are rotating into the available open
space of the lattice, the greatest possible fractional contraction 
corresponds to the condition of corner-sharing octahedra,
$\tilde{r}_2=0$.  
However, even if $\tilde{r}_2 \approx 1$, substantial
contraction is accessible.
Because the slopes of the curves in Fig.~\ref{figure:3dvolume} 
are vanishingly small for small octahedral rotations, lattice
contraction is only possible if the rotations are sufficiently
large enough to compensate for any expansion of the octahedra.

The present model in the limit of corner-shared octahedra
($\tilde{r}_2=0$) is identical to the structure of ReO$_3$.
The discussion above suggests that the RUM model for
NTE in ReO$_3$ requires the 
rotations of the octahedra to be large enough to 
compensate for any expansion of the octahedral length.  
(Although in the present simple rigid mechanical
model cubic symmetry is broken, it must be true
that any rigid unit rotation causes volume
contraction proportional to the cosine of some 
angle of rotation, which will have a similar 
vanishingly small effect for small rotations.)
The changeover to positive thermal
expansion (PTE) is understandable as being a result of octahedral
expansion overwhelming the contracting effect of rotations.
If octahedral expansions have begun to overwhelm 
the effect of rotations at 220~K, then 
what might cause the observed return to NTE
in the 600~K to 680~K range?
More generally, we ask if there are microscopic material
properties that might be manipulated to control the
magnitude and temperature range of NTE?
While the rigid mechanical model presented
thus far has provided some insight, 
the dynamics of these simple systems is necessary
to gain a more complete understanding.

\subsection{Classical molecular dynamics}
\label{subsection:dynamics}

Even though octahedra are relatively stiff, 
they still expand and/or distort as a result of
thermal motions.  In order to assess the effects of
these distortions on the volume of the crystal,
we implemented molecular dynamics simulations for the
present three-dimensional model using 
\textsc{lammps}.\cite{Plimpton95p1}
Interactions between atoms were modeled with Morse potentials,
with energy given by 
\begin{equation}
\label{eq:morse}
E_\mathrm{Morse}(r) = D\left[ e^{-2\alpha (r-r_0) }
- 2 e^{-\alpha (r-r_0)} \right]\, ,
\end{equation}
where $D$ is the dissociation energy, 
$\alpha$ controls the width of the well, and
$r_0$ is the zero temperature equilibrium distance. 
The parameters used in this work are displayed 
in Table~\ref{table:morseparameters}.
We employ strong Morse bonds 
between the central $A$ atom of the octahedra and the $X$ atoms 
forming the octahedral vertices.  To provide controllable 
rigidity of the octahedra, we also include Morse interactions
between each $X$ atom and its four nearest-neighbor $X$
atoms on the same octahedron.
We performed two separate sets of calculations with 
different arrangements of second-neighbor interactions.
In the first case, we include only second-neighbor interactions
between specific pairs of second-neighbor $X$ atoms corresponding
to the mechanical links that we described in 
Section~\ref{subsection:mechanical}.
In the second case, we include interactions between each $X$
atom and all six of its second nearest-neighbor $X$ atoms in
a lattice of corner-connected octahedra.
\begin{table}
\caption{\label{table:morseparameters}Morse potential
parameters used in molecular dynamics simulations presented
in this work.}
\begin{ruledtabular}
\begin{tabular}{lccc}
Bond & $D$ (eV) & $\alpha$ (\AA$^{-1}$) & $r_0$ (\AA)\\
\hline
\multicolumn{4}{c}{Non-corner-connected rigid octahedra} \\
\hline
$A$-$X$ & 5.00 & 5.53 & 2.00 \\
$X$-$X$ ($1^\mathrm{st}$ neighbor-onsite\footnotemark[1]) & 5.00 & 5.53 & 2.83 \\
$X$-$X$ ($1^\mathrm{st}$ neighbor-offsite\footnotemark[2]) & 5.00 & 5.53 & 1.50 \\
$X$-$X$ ($2^\mathrm{nd}$ neighbor) & 5.00 & 5.00 & 5.70 \\
\hline
\multicolumn{4}{c}{Corner-connected rigid octahedra}\\
\hline
$A$-$X$ & 5.00 & 5.53 & 2.00 \\
$X$-$X$ ($1^\mathrm{st}$ neighbor) & 5.00 & 5.53 & 2.83 \\
$X$-$X$ ($2^\mathrm{nd}$ neighbor) & 5.00 & 0.50 & 4.00 \\
\hline
\multicolumn{4}{c}{Corner-connected flexible octahedra}\\
\hline
$A$-$X$ & 5.00 & 5.53 & 2.00 \\
$X$-$X$ ($1^\mathrm{st}$ neighbor) & 0.30 & 0.50 & 2.83 \\
$X$-$X$ ($2^\mathrm{nd}$ neighbor) & 0.30 & 0.50 & 4.00 \\
\end{tabular}
\end{ruledtabular}
\footnotemark[1]{$X$ atoms on the same octahedron.}
\footnotemark[2]{$X$ atoms on different octahedra.}
\end{table}

Simulations, at fixed number, temperature, and pressure,
were run for $13$ ns (6.5 million steps) 
in supercells consisting of $19,208$ atoms and $65,856$ bonds.
In all cases, the starting configuration was at the
the global minimum of the potential energy, 
corresponding to the maximum volume arrangement, with 
octahedral vertices aligned with the cubic supercell 
lattice directions.  Average values of the lattice 
parameters were computed by sampling every $10,000$ steps
after the first two million steps.

\subsubsection{Tetragonal second-neighbor bond arrangement}

\begin{figure}
\includegraphics[width=3in]{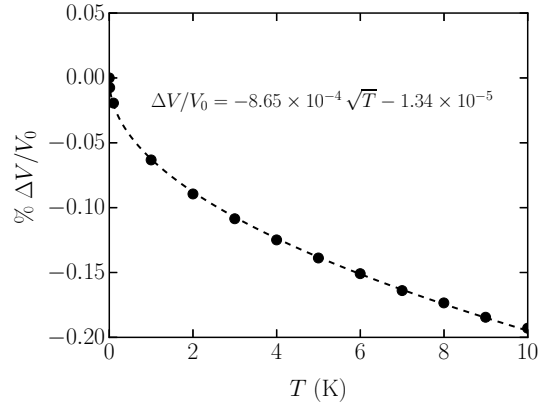}%
\caption{\label{figure:nte}
The volume decreases from the reference volume as the square root
of the temperature $T$.  The dashed line represents
the fit of the data to the square root of the temperature.
Above the temperature displayed, thermal motions drive the
system into a much lower volume as a result of an unstable
arrangement of the pattern of second-neighbor bonds. 
We note that zero-point motion would be important 
in an accurate treatment at these temperatures. }
\end{figure}

\begin{figure}
\includegraphics[width=3in]{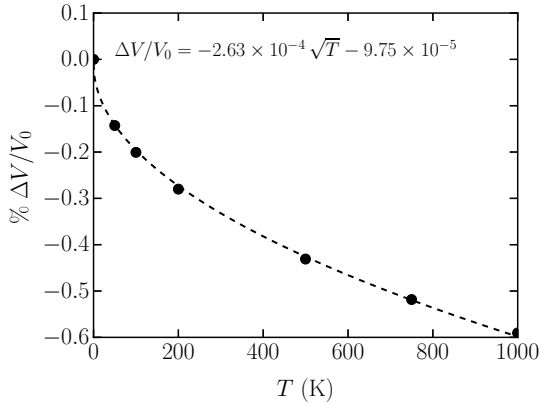}%
\caption{\label{figure:ntestable} 
Second-neighbor bond arrangement was updated to ensure 
lattice stability.  Again, the volume 
decreases from the reference volume as the square root
of the temperature $T$, however the magnitude of expansion
is somewhat less than that displayed in Fig.~\ref{figure:nte}.  
The dashed line is a fit of the data to the square root 
of the temperature.  We note that the magnitude of NTE is nearly
3.3 times smaller than it is for the unstable lattice 
(see Fig.~\ref{figure:nte}).
}
\end{figure}

Simulations for octahedra without shared corners readily exhibited NTE.  
(See the examples in Figs.~\ref{figure:nte} and \ref{figure:ntestable}.)
The volume change with $T$ is found to be proportional to $\sqrt{T}$.
We studied arrangements of the second-neighbor interactions
in our models.  In one arrangement, there
remained a near-zero-energy pathway for the octahedra to rotate,
which led to a collapse of the lattice to a low volume
configuration.
At low enough temperatures, the
collapse is avoided and NTE is observed;
above the collapse temperature, NTE is still found.
The collapse is a result of 
vibrationally-induced coherent shifts of layers of octahedra 
in an alternating pattern along a $\langle 100 \rangle $ 
direction in the lattice.  Above the collapse, the NTE coefficient
is lower.
In the second arrangement of second-neighbor bonds, 
the instability is removed and NTE is retained, with the supercell
contracting tetragonally, similar to the rigid mechanical
modeling discussed above.  However, although the number of bonds
and their parameters are the same for both the unstable and stable
lattices, the magnitude of NTE of the stable lattice is smaller than
that of the unstable lattice by a factor of $\sim 3.3$.  
Evidently, the instability assists NTE.

The positions of the octahedral vertex $X$-atoms
are distributed more broadly perpendicular to the $A$-$X$ bonds,
with transverse mean-squared displacements about five times larger
than the mean-squared displacements along the bond.  
Furthermore, the transverse mean-squared displacements also
display a temperature dependence proportional to $\sqrt{T}$.
Angular deviations of the bonds grow with
temperature, deviating by a mean of about $2^\circ$
and a standard deviation of about $1^\circ$ at the
highest temperature shown in 
Fig.~\ref{figure:ntestable}, consistent with larger
motions of the vertex ions in the directions transverse to the
bonds to central atoms.

\begin{figure}
\includegraphics[width=3in]{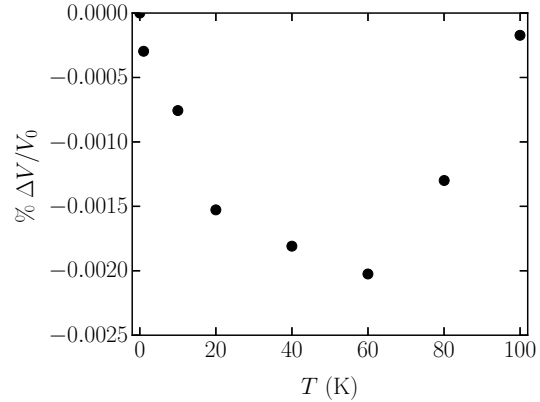}%
\caption{\label{figure:sharedcorners}
Percentage relative volume change is plotted 
as a function of temperature $T$ using the Morse
parameters for corner-connected rigid octahedra (see
Table~\ref{table:morseparameters}). 
The lattice of corner-sharing rigid octahedra results in a 
thermal expansion curve moving from (weak) NTE to PTE,
a distribution typical of many NTE materials.  Further
calculations at higher $T$ resulted in strong PTE.}
\end{figure}

A lattice of corner-sharing octahedra was also constructed, 
corresponding to the structure of ReO$_3$.  Also included were 
selected second-neighbor interactions imposing tetragonal symmetry.  
Simply transferring the Morse potential parameters that 
produced NTE in the case of separated octahedra resulted in 
positive thermal expansion (PTE), with the volume linearly
increasing with temperature; the octahedral expansion
clearly had overwhelmed any contraction due to rotations.  
We changed the Morse parameters of the second-neighbor interaction, 
reducing the restoring force to $\approx 1\%$ of the 
first-neighbor restoring force.  Subsequent
simulations produced weak NTE that gives way to
PTE as the temperature is increased  
(Fig.~\ref{figure:sharedcorners}).  This behavior is strikingly
similar to the low-temperature thermal expansion in many materials
that exhibit NTE, especially the case of 
ReO$_3$.\cite{Chatterji09p241902}

The volume contraction behavior ($\propto \sqrt{T}$) 
indicates a relationship to the typical speeds of the
particle motions or, equivalently, to the typical
mean-squared displacements of the atoms bound in a
Morse potential.  We also note the above observation 
that the velocities of vertex atoms are biased into 
directions perpendicular to their bonds with the
central atoms of the octahedra.  The bond-expansion
inherent in the Morse potential provides the force
that drives atoms farther apart.\cite{supplemental} 
In the present model, the nearest-neighbor coordinations,
combined with weak second-neighbor
interactions, guide the vertex atoms into the 
open spaces of the lattice, resulting in the observed NTE.  
In shared-corner octahedral structures, the transverse 
motions become overwhelmed by the 
expansion of bonds to the central atoms at sufficiently
high temperatures, leading to the eventual change-over to 
positive thermal expansion.  Adding links that separate
the octahedra enables stronger NTE and pushes to 
much higher $T$ the
eventual changeover to positive thermal expansion.

\subsubsection{Isotropic second-neighbor interactions}

\begin{figure}
\includegraphics[width=3in]{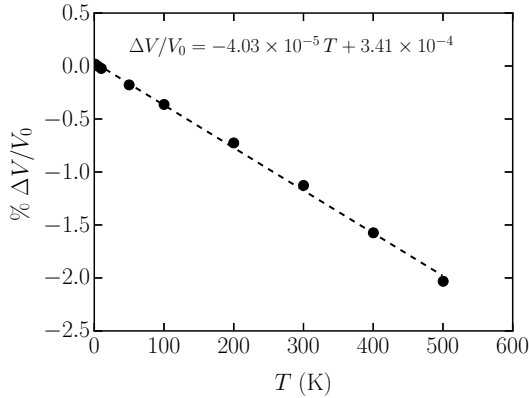}%
\caption{\label{figure:isotropic-nte}
Percentage relative volume change is plotted 
as a function of temperature corner-connected flexible 
octahedra.  All vertex atoms interact with all six second-neighbor
vertex atoms.  Weak first- and second-neighbor 
interactions produce substantial NTE over a wide temperature 
range. A straight line is fit to the data.}
\end{figure}

\begin{figure}
\includegraphics[width=3in]{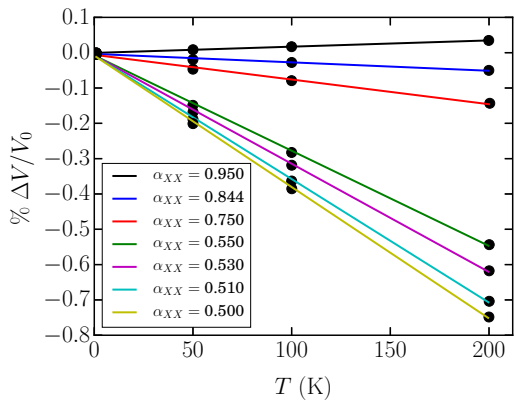}%
\caption{\label{figure:isotropic-many}
(Color online) Percentage relative volume change is plotted as a function
of temperature for several $X$-$X$ interaction strengths, as
controlled by the Morse $\alpha$ parameter.
Thermal expansion is shown to vary from NTE to PTE as
the strength of the $X$-$X$ interactions increase.
In all these plots, the well depths are held fixed at the
values used for corner-connected flexible octahedra.
}
\end{figure}

When all interactions between $X$-atoms and their six
second-neighbor vertex $X$-atoms for corner-connected 
octahedra are included, eliminating the artificially 
imposed tetragonal symmetry, NTE is still observed.  
In this arrangement, lattice contraction 
is only possible when the strength
of the $X$-$X$ nearest-neighbor interaction is
reduced to the same magnitude as the second-neighbor
interactions, producing readily deformable octahedra
and compromising the stability of the lattice.
The NTE observed is isotropic, has a strong magnitude, 
and exists over wide temperature range up to the point
where the material begins to melt 
(see Fig.~\ref{figure:isotropic-nte}).
We observe that NTE is linear in temperature, contrasting
with the behavior for the results displayed above.
The mean-squared displacements of
the vertex atoms transverse to their nearest-neighbor
bonds expand \emph{linearly} in temperature,
contrasting with the square-root dependence 
observed for rigid octahedra.
The linearity of NTE in temperature is a result
of the octahedra being much more flexible to angular
distortions than in the examples above.  Linearity
is also a characteristic of the thermal expansion
of a two-atom system bonded in the Morse potential.
%
%
Varying the strength of the $X$-$X$ interactions,
by adjusting the Morse $\alpha$ (see Eq.~\ref{eq:morse})
permits control of thermal expansion in the model 
from NTE to PTE (see Fig.~\ref{figure:isotropic-many}).

\section{Discussion}

The common feature across all the materials exhibiting
NTE is an open lattice.  However, the fact that different
materials with the same lattice possess different thermal
expansion characteristics demonstrates the importance
of bond strength ratios to thermal expansion.
Bonding is essential in determining the equilibrium
structure of a material.  
In addition, when there are competing phases with
low enough energy differences, states with higher
entropy will be favored as temperature increases.
The present model calculations, in which the
structural stability is maintained by 
second-neighbor interactions, 
demonstrate that reducing the strength of
second-neighbor interactions leads to NTE, 
which possibly is due to increased accessibility
to competing phases.
Materials with large unit cells, such as
ZrW$_2$O$_8$ and Li$_2$B$_4$O$_7$ readily provide support
for this picture.
ZrW$_2$O$_8$ is known to be metastable\cite{Arora04p1025}
and metastability such as this has been cited as a likely
underlying reason for NTE.\cite{Liu11p664}
It is furthermore noted that synthesis 
of good Li$_2$B$_4$O$_7$ samples is hampered 
by the existence of a number of (meta-)stable 
phases under growth conditions.\cite{Senyshyn10p093524}

In contrast to these complex materials, which have competing metastable 
states and provide one view of NTE, we point out that
NTE is observed even in the simplest materials, such as
those with diamond or other low-coordination lattices.
In early lattice dynamics modeling with spherically 
symmetric potentials, the diamond lattice was observed 
to exhibit NTE.\cite{Wallace65pA152}  We reproduced
this observation in our own molecular dynamics calculations.
Furthermore, we found that by increasing the strength of 
second-neighbor interactions, stabilizing the lattice, we can
eliminate NTE in this lattice as we did for corner-connected octahedral
model discussed above.  Experimentally, NTE in Ag$_2$O has a greater
magnitude than in Cu$_2$O, while it is nearly nonexistent in
Au$_2$O, and this magnitude was observed to correlate, in DFT 
calculations, with the degree to which the charge density
is spherically distributed,\cite{Gupta14p093507} which again
supports the view that lower directionality of bonds is important.
In all of these examples as in our model calculation, 
the open lattice and the weak second-neighbor bond potentials,
implying highly flexible bond angles, are 
the essential ingredients that yield NTE.  The reasonable 
success of the RUM in describing NTE is a result 
of the fact that there are strong bonds within the polyhedra,
minimizing their expansion, coupled with low resistance
to rotations.
We observe from the current calculations that reducing 
the resistance to changing bond angles within polyhedra 
yields stronger NTE behavior with a constant coefficient of
expansion, such as that seen in ZrW$_2$O$_8$, lending further
support to previous findings\cite{Sanson14p3716} that
models involving coherent motions of atoms, such as the RUM,
are not consistent with the NTE in this material.
Finally, from the observation in the current model that 
angular deviations within the octahedra is a key to the 
emergence of NTE leads us to suggest that the observed 
reappearance of NTE in ReO$_3$ at higher temperature 
could be a result of temperature-induced
reduced rigidity of the octahedra 
permitting more atomic motions to expand into 
remaining lower-density regions of the lattice.  
However, confirmation of this is beyond the scope of
the present work.

\section{Conclusion}

We constructed model systems
consisting of point masses joined by anharmonic springs,
analytically showed that the RUM is a decent description
of NTE as long as coherent rotations of the rigid units
is sufficient to outweigh any uniform expansion of the
same rigid units.  We showed that NTE can be strongly
enhanced by increasing the flexibility of previously considered
to be rigid units.  Finally, we demonstrated that a merely modest
reduction in this flexibility is enough to restore positive expansion.


\begin{acknowledgments}
The authors acknowledge the support of the Department of Energy, Office 
of Basic Energy Sciences, through grant DE-FG02-07ER46431. Computational 
support was provided by the National Energy Research Scientific 
Computing (NERSC) Center.%
\end{acknowledgments}

%

\end{document}



\title{Classical model of negative thermal expansion in solids
due to positively expanding bonds: Supplementary material}


\author{Joseph T. Schick}
\affiliation{Department of Physics, Villanova University, Villanova, 
PA 19085, USA}

\author{Andrew M. Rappe}
\affiliation{The Makineni Theoretical Laboratories, Department of Chemistry, 
University of Pennsylvania, Philadelphia, Pennsylvania 19104-6323, USA}


\date{\today}


\pacs{}

\maketitle


The Morse potential\cite{Morse29p57}
captures the essential anharmonicity necessary for
describing the connection between bond expansion and
thermal expansion properties of materials.
Furthermore, the Morse potential
has an easily calculable thermal expansion which makes it
especially suitable for this model calculation.
The Morse potential between two atoms separated by $r$ 
is given by the expression
\begin{equation}
\label{eq:Morse}
E_\mathrm{Morse}(r) = D \left[ e^{-2\alpha ( r - r_0)} - 
2e^{-\alpha(r - r_0)} \right] \, ,
\end{equation}
where $r_0$ is the equilibrium separation, $D$ is the dissociation energy
and $\alpha$, with dimension of inverse length, controls the width
of the potential well.


\begin{figure}
\includegraphics[width=3.2in]{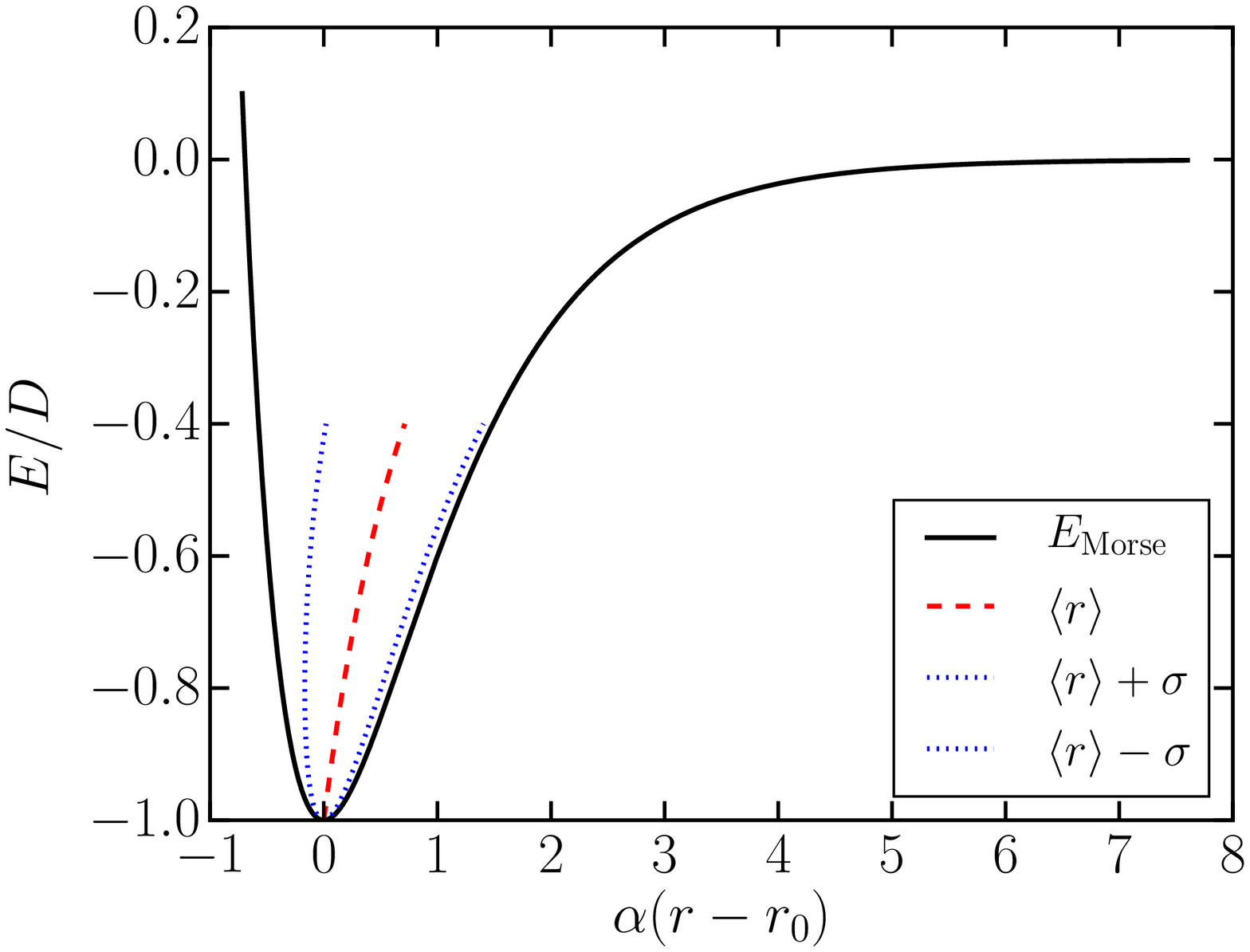}%
\caption{\label{figure:morse}The expansion of 
the one dimensional Morse oscillator
is displayed in terms of energy normalized to $D$, which is
the dissociation energy.  The mean value of the displacement (red dashes) 
is evaluated analytically (Eq.~\ref{eq:expansion}) and the root-mean-squared
deviations of the displacement (blue dots) 
are computed numerically from the equation of motion.
Turning points of classical motion (black curves) are displayed for reference.}
\end{figure}

The solution of the equation of motion of the 
classical Morse oscillator for bounded motion has a simple
analytic form, (\emph{e.g.} see Ref.~\onlinecite{DeMarcus78p733}),
\begin{equation}
\label{eq:morsemotion}
r(t) - r_0 = \frac{1}{\alpha}
\ln \left[
\frac{1-\sqrt{\epsilon} \cos \omega t}
{1-\epsilon}
\right];~\quad (\epsilon < 1)\, ,
\end{equation}
where $\epsilon = (E_\mathrm{Morse}+D)/D$ is the energy of the oscillator
as a fraction of the dissociation energy $D$ and
the angular frequency $\omega$ is given in terms of the
potential parameters and the reduced mass $\mu$ by
\begin{equation}
\omega^2 = \frac{2\alpha^2 D}{\mu}\left( 1 - \epsilon \right)\, .
\end{equation}
Integrating Eq.~\ref{eq:morsemotion} over one period, 
we determine the mean displacement
\begin{equation}
\label{eq:expansion}
\langle r - r_0 \rangle
= \frac{1}{\alpha} \ln\left[ 
\frac{1+\sqrt{1-\epsilon}}{2(1-\epsilon)} \right]\, .
\end{equation}
We plot this value along with numerically determined
root-mean-squared deviations in Fig.~\ref{figure:morse}.

We are interested in the low energy limit, at which
Eq.~\ref{eq:expansion} is found to be linear in energy,
\begin{equation}
\langle r - r_0 \rangle \approx \frac{3\epsilon}{4\alpha}\, .
\end{equation}
To obtain the low energy behavior of the mean-squared displacement,
$\langle ( r - r_0 )^2 \rangle$,
we expand Eq.~\ref{eq:morsemotion}
to sixth order in $\sqrt{\epsilon}$ and then integrate to find
\begin{equation}
\label{eq:msd}
\langle ( r - r_0 )^2 \rangle = \frac{1}{\alpha^2} \left[ \frac{1}{2}\epsilon
+ \frac{27}{32}\epsilon^2 + O(\epsilon^3)\right]\, ,
\end{equation}
implying the dimensionless root-mean-squared displacement is
$\sigma = \sqrt{\epsilon/2}$ to lowest order in $\epsilon$.